\documentclass{sag00}
\draft

\renewcommand{\arraystretch}{1.5}
\begin{document}
\SetRunningHead{Author(s) in page-head}{Running Head}
%\Received{}%{yyyy/mm/dd}
%\Accepted{}%{yyyy/mm/dd}
%\Published{}%{yyyy/mm/dd}

%\Jtitle{論文のタイトル}
\title{Detection of Intermediate-Mass Black Holes
in Globular Clusters Using Gravitational Lensing}

%%% begin:list of Jauthors
% Do NOT capitalize all letters in "textsc".
%\Jauthor{A-Firstname \textsc{A-Familyname} %
%  \thanks{Example: Present Address is xxxxxxxxxx}}
%\Jaffil{A-Address of Institute}

%\Jauthor{B-Firstname \textsc{B-Familyname}}
%\Jaffil{B-Address of Institute}
%\and
%\Jauthor{C-Firstname {\sc C-Familyname}}
%\Jaffil{C-Address of Institute}
%%% end:list of authors

%%% Please use the following style in case that sorting by 
%%% affiliation is impossible. 
%
% \Jauthor{%
%   D-Firstname \textsc{D-Familyname}\Jaltaffilmark{1}
%   E-Firstname \textsc{E-Familyname}\Jaltaffilmark{1,2}
%   and
%   F-Firstname \textsc{F-Familyname}\Jaltaffilmark{2}}
% \Jaltaffiltext{1}{Address of Institute}
% \Jaltaffiltext{2}{Address of Institute}

%%% begin:list of authors
% Do NOT capitalize all letters in "textsc".
\author{Takayuki \textsc{Tatekawa} %
           }
\affil{Department of Social Design Engineering, National Institute of Technology,
Kochi College,\\
200-1 Monobe Otsu, Nankoku, Kochi, 783-8508, Japan \\
Research Institute for Science and Engineering, Waseda University,\\
3-4-1 Okubo, Shinjuku, Tokyo 169-8555, Japan}
\email{tatekawa@kochi-ct.ac.jp}
\and
\author{Yuuki \textsc{Okamura}}
\affil{Department of Electrical Engineering and Information Science, National Institute of Technology,
Kochi College,\\
200-1 Monobe Otsu, Nankoku, Kochi, 783-8508, Japan}%\email{bbbbb@xxx.xxx.xx.xx}
%\and
%\author{C-Firstname {\sc C-Familyname}}
%\affil{C-Address of Institute}\email{ccccc@xxx.xxx.xx.xx}
%%% end:list of authors

%%% Please use the following style in case that sorting by 
%%% affiliation is impossible. 
%
% \author{%
%   D-Firstname \textsc{D-Familyname}\altaffilmark{1}
%   E-Firstname \textsc{E-Familyname}\altaffilmark{1,2}
%   and
%   F-Firstname \textsc{F-Familyname}\altaffilmark{2}}
% \altaffiltext{1}{Address of Institute}
% \email{ddddd@xxx.xxx.xx.xx}
% \email{eeeee@xxx.xxx.xx.xx}
% \altaffiltext{2}{Address of Institute}

%% `\KeyWords{}' always has to be placed before `\maketitle'.
\KeyWords{Gravitational lensing: IMBH: globular clusters} %Do NOT move this preamble from here!

\maketitle

%\begin{Jabstract}
%原稿を準備する前に、``IMPORTANT NOTICE'' を注意深くお読みください。
%\end{Jabstract}

\begin{abstract}
Recent observations suggest the presence of supermassive black holes
at the centers of many galaxies. The existence of intermediate-mass black holes (IMBHs) in globular clusters has also been predicted. We focus 
on gravitational lensing as a new way to explore these entities. It is known that the
mass distribution of a self-gravitating system such as a
globular cluster changes greatly depending on the
presence or absence of a central massive object.
After considering possible mass distributions for a globular cluster belonging to the Milky Way galaxy, we estimate that the effect on the separation
angle of gravitational lensing due to an IMBH would be of milliarcsecond order.
\end{abstract}

%\section{Introduction}
%
%\noindent IMPORTANT NOTICE\\
%1. ``\verb|\draft|'' creates single column and double spaces format.\\
%2. If you comment out ``\verb|\draft|'', the output will be double column
%   and single space.\\
%3. For cross-references, the use of ``\verb|\label|, \verb|\ref|, \verb|\cite|'' 
%   and the thebibliography environment is strongly recommended. \\
%4. Do NOT use ``\verb|\def|, \verb|\renewcommand|''.\\
%5. Do NOT redefine commands provided by SAG00.cls.\\

%%%%%%%%%%%%%%%%%%%%%%%%%%%%%%%%%%%%%%%%
%%% Introduction
%%%%%%%%%%%%%%%%%%%%%%%%%%%%%%%%%%%%%%%%

\section{Introduction}	
The existence of supermassive black holes (SMBHs) at the centers
of galaxies has been made evident by recent observations.
For example, the shadow of the SMBH in the center of M87 was directly
observed by the Event Horizon Telescope (EHT)~\citep{EHT}.
Long-term observations of the movement
of stars surrounding the center of the Milky Way Galaxy suggest the existence of an
invisible massive object, Sgr A${}^*$,
considered to be an SMBH~\citep{Ghez2000,Gillessen2009}.
For other galaxies, the existence of SMBHs is indirectly
suggested by the $M-\sigma$ relation between the mass of the SMBH and the velocity dispersion
of stars in a galaxy~\citep{Silk1998,Ferrarese2000,Gebhardt2000,Gultekin2009}.

The question of how SMBHs form has not yet been definitively answered, 
although various scenarios have been considered over the years
~\citep{Rees1984}. One of the scenarios involves an 
intermediate-mass black hole (IMBH)
~\citep{Greene2020} that grows to become an SMBH.
On the basis of the $M-\sigma$ relation, it seems possible that IMBHs may be located at the centers
of globular clusters. This has been discussed particularly
in the case of M15 but has not been resolved
~\citep{Gerssen2002,Baumgardt2003,McNamara2003,Kiselev2008,Murphy2011,denBrok2014},
although indirect verification has been performed
using statistical properties of stars.

In this paper, we consider a new method of detecting
SMBHs and IMBHs. These celestial objects' immense mass
bends the trajectory of light from background objects: 
the gravitational lensing (GL) effect. It can cause background objects to be observed multiple times,
and single images to be brightened. Through the GL effect, it is possible to
verify the existence of a massive object between the background light-source and the Earth.

GL by globular clusters has been discussed in the past.
Kains et al.~(2016,2018) proposed a method for IMBH detection
using gravitational microlensing.
Bukhmastova~(2003) has attempted to explain
QSO-galaxy associations using the GL effect
produced by globular clusters.

GL is a small effect even for SMBHs;
our treatment includes lensing by both the massive object and the surrounding stars. The density profile of the
stars in the cluster may be obtained from the equilibrium solution for 
a gravitational many-body system; it varies
greatly depending on the presence or absence of
a massive central object. In this paper, we assume a spherically
symmetric distribution of stars and analyze the GL effect for various
characteristic density profiles.

The paper is organized as follows. In Section 2, we discuss
the mass distribution for globular clusters.
Two models of the distribution are described: 1) a model based on the static state of
self-gravitating systems and 2) a phenomenological model.
If there is an IMBH in a globular cluster, the surrounding stars 
are thought to form a cusp. Therefore,
the two models are also evaluated with such a cusp, so that in all
we consider four types of models in this paper.
In Section 3, the geodesic equation for light trajectories is discussed.
Since the models assume spherical symmetry,
the spacetime can be described by the Schwarzschild metric.
We consider the gravitational potential of each model in this metric.
If the cusp is distributed across the entire area, the mass will diverge.
Hence, it is necessary to connect the spacetime inside the cusp smoothly with that of the cuspless model.
In Section 4, we compare the GL
effect in mass distribution models with/without an IMBH, and
estimate the maximum separation angle caused
by GL.
The difference in the separation angle is found to be of
sub-milliarcsecond order when we assume that the lensing object is
a globular cluster in the Milky Way Galaxy.
In Section 5, the conclusions of this study are presented.

%%%%%%%%%%%%%%%%%%%%%%%%%%%%%%%%%%%%%%%%
%%% Mass Distribution
%%%%%%%%%%%%%%%%%%%%%%%%%%%%%%%%%%%%%%%%

\section{Mass Distribution}
\subsection{Mass Distribution for Globular Clusters}
In this subsection, we explain mass distributions for globular clusters.
For simplicity, we assume a spherically symmetric distribution and consider the equilibrium solution
in a self-gravitating system under Newtonian gravity. Although the distribution
function in general depends on both energy and angular momentum,
we assume that the effect of angular momentum is small and only the energy dependence 
need to be considered. In this case, the velocity dispersion is isotropic at each
point in space, and the energy for matter is given by the kinetic
and potential energies. 

The distribution function tells us both the distribution of the matter in the system
and the velocity at which it moves. The stars in the cluster may be thought of
as a fluid obeying an equation of continuity, the collisionless Boltzmann equation, 
that contains a term depending on the gradient of a potential. Suppose that the
distribution function is a known function of energy, and that the potential in 
the collisionless Boltzmann equation is the gravitational potential solving the
Poisson equation. The mass density is of course proportional to the gradient 
of this potential. Therefore, it is possible in principle (if not always in practice)
to solve the collisionless Boltzmann equation and Poisson's equation together
to obtain the mass density. In the case that the distribution function goes as 
a power of the energy, the combined equation is called the Lane-Emden equation; 
the density is proportional to solution of this equation raised to the power
of the polytropic index $n$. Unfortunately, the coefficient of the proportionality
is not a constant, but a generally complicated function. Only in the case of 
$n=5$, the Plummer model \citep{Plummer,GalacticDynamics}, does the
Lane-Emden equation yield a simple expression for the density.
\begin{equation} \label{eqn:rhoP}
\rho_P(r) = \frac{3}{4\pi} \frac{M_{\rm tot} r_0^2}{(r^2 + r_0^2)^{5/2}} \,,
\end{equation}
where $r_0$ represents ``Plummer length''.
$M_{{\rm tot}}$ is the total mass of the cluster.

Another well-known model is that of Hernquist, which realizes 
de Vaucouleurs’ $1/4$ law, the relationship between surface brightness
and distance from the center for elliptic galaxies~\citep{deVaucouleurs}. 
Although the Hernquist model is phenomenological, it has been widely applied,
and with much success~\citep{Hernquist}. The density in the Hernquist model
is given by:
\begin{equation} \label{eqn:rhoH}
\rho_H (r) = \frac{M_{\rm tot}}{2 \pi} \frac{r_0}{r} \frac{1}{(r+r_0)^3} \,.
\end{equation}
Note that the Hernquist model diverges at $r=0$, whereas in the Plummer model, the density distribution converges gently at the center. 
  
For a globular cluster without an IMBH, we consider the Plummer model
and the Hernquist model. 

\subsection{Mass Distribution Around an IMBH}
If a globular cluster includes an IMBH, the exchange of  
orbital energies causes the distribution
of stars to take a characteristic form. Here we make several
assumptions: 1) The distribution of stars is
represented by a single-particle distribution function; 
2) The mass of the IMBH
is much smaller than the mass of the globular cluster core;
3) For simplicity, all the stars around the IMBH have the same mass;
4) The distribution of stars is independent of
angular momentum.
Under these assumptions, the static solution for
the density distribution of stars by the Fokker-Planck
equation obeys a specific power law and is known as the 
Bahcall-Wolf cusp~\citep{Bahcall-Wolf,Merritt-Book}.
\begin{equation}
\rho_B (r) \propto r^{-7/4}  \,.
\end{equation}
The Bahcall-Wolf cusp has been verified by $N$-body simulation
for stellar systems around a massive object~\citep{Preto2004}.
In contrast to the Plummer models, in a Bahcall-Wolf cusp
the mass density diverges at the center.
To consider lensing by a Bahcall-Wolf cusp,
it is necessary to consider the mass of the IMBH itself.

%%%%%%%%%%%%%%%%%%%%%%%%%%%%%%%%%%%%%%%%
%%% Geodesic equations
%%%%%%%%%%%%%%%%%%%%%%%%%%%%%%%%%%%%%%%%

\section{Geodesic Equations}
\subsection{Geodesic Equations for the Spherically Symmetric Model}
In this paper, we consider the equilibrium state for globular clusters
with spherical symmetry. Therefore, spacetime is described
by the Schwarzschild metric~\citep{Gravitation,Hartle}.
\begin{equation}
{\rm d} s^2 = - \left (1+\frac{2}{c^2} \Psi(r) \right )
 {\rm d} (ct)^2 + \left (1+\frac{2}{c^2} \Psi(r) \right )^{-1}
 {\rm d} r^2 + r^2 \left ( {\rm d} \theta^2 
 + \sin^2 \theta {\rm d} \phi^2 \right ) \,,
\end{equation}
where $\Psi(r)$ corresponds to the Newtonian gravitational potential.

From the Schwarzschild metric, geodesic equations may be derived.
Hereafter we define the time component as
\begin{equation}
w \equiv ct \,.
\end{equation}

The geodesic equations are as follows:
\begin{eqnarray}
\frac{{\rm d}^2 w}{{\rm d} \lambda^2}
 &=& -\frac{2}{c^2} \frac{{\rm d} \Psi}{{\rm d} r}
 \left ( 1+ \frac{2}{c^2} \Psi \right ) \frac{{\rm d} w}{{\rm d} \lambda}
 \frac{{\rm d} r}{{\rm d} \lambda} \,, \\
\frac{{\rm d}^2 r}{{\rm d} \lambda^2}
 &=& -\frac{2}{c^2} \frac{{\rm d} \Psi}{{\rm d} r}
 \left ( 1+ \frac{2}{c^2} \Psi \right )
  \left( \frac{{\rm d} w}{{\rm d} \lambda} \right )^2
 + \frac{2}{c^2} \frac{{\rm d} \Psi}{{\rm d} r}
 \left ( 1+ \frac{2}{c^2} \Psi \right )^{-1}
  \left( \frac{{\rm d} r}{{\rm d} \lambda} \right )^2 \nonumber \\
 && + r \left ( 1+ \frac{2}{c^2} \Psi \right )
  \left( \frac{{\rm d} \vartheta}{{\rm d} \lambda} \right )^2
  + r \sin^2 \vartheta
  \left ( 1+ \frac{2}{c^2} \Psi \right )
  \left( \frac{{\rm d} \phi}{{\rm d} \lambda} \right )^2 \,, \\
\frac{{\rm d}^2 \vartheta}{{\rm d} \lambda^2}
 &=& -\frac{2}{r} \frac{{\rm d} r}{{\rm d} \lambda}
  \frac{{\rm d} \vartheta}{{\rm d} \lambda}
 + \sin \vartheta \cos \vartheta 
  \left( \frac{{\rm d} \phi}{{\rm d} \lambda} \right )^2 \,, \\
\frac{{\rm d}^2 \phi}{{\rm d} \lambda^2}
 &=& -\frac{2}{r} \frac{{\rm d} r}{{\rm d} \lambda}
  \frac{{\rm d} \phi}{{\rm d} \lambda}
 - \frac{2 \cos \vartheta}{\sin \vartheta}
  \frac{{\rm d} \vartheta}{{\rm d} \lambda} 
  \frac{{\rm d} \phi}{{\rm d} \lambda} \,.
\end{eqnarray}
Because we consider a null geodesic, we have introduced the parameter $\lambda$
instead of the world interval $s$. 
Moreover, because the trajectory of light can be modeled in a plane, 
when we notice one trajectory of light, we can ignore the
$\phi$ components. Hereafter we set $\vartheta=\pi/2$.

\subsection{Derivation of Potential Term for Models}
We consider four models for globular clusters. When there is no IMBH,
the densities given by Eqs. (\ref{eqn:rhoP})
and (\ref{eqn:rhoH})correspond to the potentials
\begin{eqnarray}
\Psi_P (r) &=& - \frac{G M_{\rm tot}}{(r^2+r_0^2)^{1/2}} \,, \\
\Psi_H (r) &=& - \frac{G M_{\rm tot}}{r+r_0} \,,
\end{eqnarray}
where $\Psi_P$ and $\Psi_H$ refer to the potentials in the
Plummer and Hernquist models, respectively.
When the globular cluster includes an IMBH, we set a Bahcall-Wolf
cusp at the center.
Outside the cusp, the effect
of the IMBH is tiny. At the boundary of the cusp $r=R$, we connect 
smoothly with the Plummer or Hernquist model. 
In connecting the models, we first focus on the continuity of the terms 
of the geodesic equation. Then we set the first derivative of
the potential to be continuous.

For the Plummer model, 
in the inner region of the model ($r<R$), the potential
is then
\begin{eqnarray}
\Psi_{\rm P+B (in)} (r) &=& -\frac{G M_{\rm BH}}{r}
 - \frac{G M_{\rm tot}}{(R^2 + r_0^2)^{1/2}}
 + \frac{64}{5} \pi G R^2 \rho_0
 \left [ \left ( \frac{r}{R} \right )^{1/4} -1 \right ] \,, \\
\frac{{\rm d} \Psi_{\rm P+B (in)} (r)}{{\rm d} r} 
 &=& \frac{G M_{\rm tot}}{r^2} + \frac{16}{5} \pi G R \rho_0
 \left ( \frac{r}{R} \right )^{-3/4} \,.
\end{eqnarray}
In the outer region of the model ($r>R$), 
the potential is
\begin{eqnarray}
\Psi_{\rm P+B (out)} (r) &=& -\frac{G M_{\rm BH}}{r}
 - \frac{G M_{\rm tot}}{(r^2 + r_0^2)^{1/2}} \,, \\
\frac{{\rm d} \Psi_{\rm P+B (out)} (r)}{{\rm d} r} 
 &=& \frac{G M_{\rm tot}}{r^2} + \frac{G M_{\rm tot} r}
 {(r^2 + r_0^2)^{3/2}} \,.
\end{eqnarray}

To satisfy the condition of connection at the boundary ($r=R$),
the density parameter $\rho_0$ must be
\begin{equation}
\rho_0 = \frac{5 M_{\rm tot}}{16 \pi (R^2+r_0^2)^{3/2}} \,.
\end{equation}

Similarly, for the Hernquist model,
in the inner region($r<R$), the potential
is
\begin{eqnarray}
\Psi_{\rm H+B (in)} (r) &=& -\frac{G M_{\rm BH}}{r}
 - \frac{G M_{\rm tot}}{R + r_0}
 + \frac{64}{5} \pi G R^2 \rho_0
 \left [ \left ( \frac{r}{R} \right )^{1/4} -1 \right ] \,, \\
\frac{{\rm d} \Psi_{\rm H+B (in)} (r)}{{\rm d} r} 
 &=& \frac{G M_{\rm tot}}{r^2} + \frac{16}{5} \pi G R \rho_0
 \left ( \frac{r}{R} \right )^{-3/4} \,.
\end{eqnarray}
In the outer region of the model ($r>R$), 
the potential is
\begin{eqnarray}
\Psi_{\rm H+B (out)} (r) &=& -\frac{G M_{\rm BH}}{r}
 - \frac{G M_{\rm tot}}{r + r_0} \,, \\
\frac{{\rm d} \Psi_{\rm H+B (out)} (r)}{{\rm d} r} 
 &=& \frac{G M_{\rm tot}}{r^2} + \frac{G M_{\rm tot} r}
 {(r + r_0)^2} \,.
\end{eqnarray}
In the Hernquist model, the density parameter $\rho_0$, which ensures
that the boundary conditions match at the point where $r=R$, 
turns out to be

\begin{equation}
\rho_0 = \frac{5 M_{\rm tot}}{16 \pi R (R+r_0)^2} \,.
\end{equation}

\subsection{Choice of Parameters}
For consideration of GL by globular clusters,
we must fix the values of some parameters. 
The mass of the
IMBH $M_{\rm BH}$, radius of the Bahcall-Wolf cusp $R$, total
mass of the lensing object $M_{\rm tot}$, and radius of the lensing
object $r_{\rm lens}$ should all be considered. 
Because, in both the Plummer and Hernquist models, a
scale length $r_0$ is included, we must decide on its value
as well.
For the null geodesic, we must consider the distance between the
center of the globular cluster and the background object $r_{\rm star}$; we must also consider
the distance $x_e$ from the Earth to the center of the globular cluster. 

Both the Plummer model and the Hernquist model include
a length parameter $r_0$ (Eqs. (\ref{eqn:rhoP})
and (\ref{eqn:rhoH})),
a characteristic quantity that determines
the shape of the density distribution.
In order to compare the two models, it is desirable to make
contained within radius $r_0$ approximately the same in both.
To do this,
we use a relation
between $r_0$ and $r_{\rm lens}$.
For the Plummer model, we set $r_0=r_{\rm lens}/2$:
\begin{equation}
M_P(r_{\rm lens}) = \int_0^{r_{\rm lens}}
 \rho_P(r) \cdot 4 \pi r^2 {\rm d} r
 = \frac{M_{\rm tot} r_{\rm lens}^3}
  {\left (r_{\rm lens}^2 + (r_{\rm lens}/2)^2
 \right )^{3/2}} \simeq 0.72 M_{\rm tot} \,.
\end{equation}
For the Hernquist model, we set $r_0=r_{\rm lens}/5$:
\begin{equation}
M_H (r_{\rm lens}) = \int_0^{r_{\rm lens}}
 \rho_H(r) \cdot 4 \pi r^2 {\rm d} r
 = \frac{M_{\rm tot} r_{\rm lens}^2}
 {(r_{\rm lens} + r_{\rm lens}/5)^2}
 \simeq 0.69 M_{\rm tot} \,.
\end{equation}
For both models, about 70 percent of total mass
is thus within the radius of the lensing object.

In this paper, the mass of the IMBH is fixed at $M_{\rm BH} = 10^4 M_{\odot}$.
The upper bound of the distance to the background star
is fixed at $r_{\rm star}<10^6 \mbox{[lyr]}$.
The globular cluster is assumed to belong to the Milky Way; catalogs of Milky Way globular clusters have been published~\citep{Harris1996,Hilker2020}.  
The parameters (total mass, 
calculated radius from apparent dimension, and distance from the Earth)
of the clusters in the present paper are chosen to be typical of relatively large
Milky Way globular clusters according to the catalogs:
\begin{itemize}
 \item $1 \times 10^6 \le M_{\rm tot} \le 5 \times 10^6 ~[M_{\odot}]$
 \item $10 \le r_{\rm lens} \le 100 ~[\mbox{lyr}]$
 \item $1 \times 10^4 \le x_e \le 10 \times 10^4 ~[\mbox{lyr}]$
\end{itemize}
Here $x_e$ means the distance from the Earth to center
of globular cluster.
The boundary of Bahcall-Wolf cusp is fixed at $R=1 [\mbox{lyr}]$.

%%%%%%%%%%%%%%%%%%%%%%%%%%%%%%%%%%%%%%%%
%%% Effect of gravitational lensing
%%%%%%%%%%%%%%%%%%%%%%%%%%%%%%%%%%%%%%%%
\section{Effect of Gravitational Lensing}
\subsection{Model Settings}
\begin{figure}
  \begin{center}
    \FigureFile(120mm,68mm){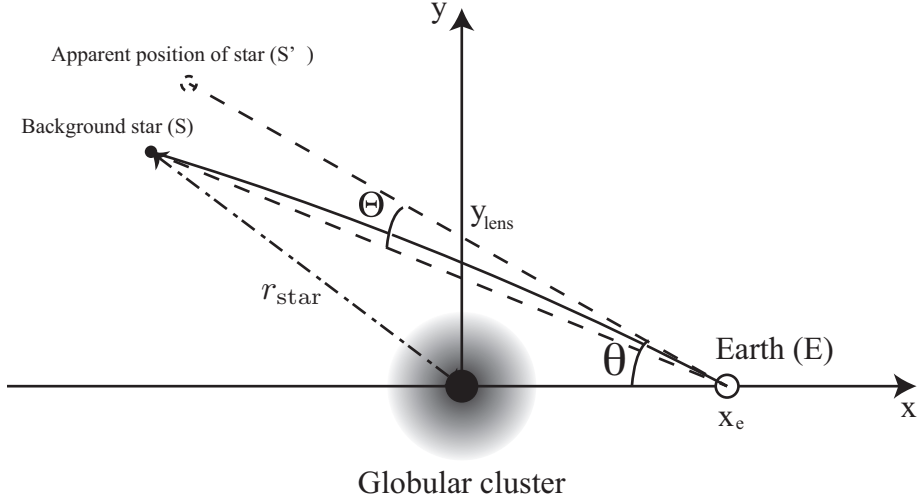}
    %%% \FigureFile(width,height){filename}
  \end{center}
  \caption{Schematic picture of the positional relationship between the Earth, globular cluster,
and background star.
The center of the globular cluster is defined as the origin, and the trajectory of light from
the background star to the Earth (solid line) is analyzed.}
\label{fig:map}
\end{figure}

Various physical and geometrical quantities required
for analyzing the GL effect of globular clusters will now be defined.
Fig.~\ref{fig:map} shows the positional relationship between the Earth, globular cluster,
and background star. 
Even if spacetime is curved by gravity, the trajectory of light can be analyzed in a plane.
The center of the globular cluster is defined as the origin. Then the Earth is set
at $(x_e, 0)$. This position is denoted as point $E$.
A background star is positioned at point $S$. Because of the effect
of GL, however, this star appears to observers on Earth to be at point $S'$.
The angle between the x-axis and line segment $\overline{ES'}$ is defined as $\theta$.
Then the deflection angle (angle between line segments $\overline{ES}$
and $\overline{ES'}$) is defined as $\Theta$.
The intersection point of the y-axis and line segment $\overline{ES'}$ is defined as the
``impact parameter'' $y_{\rm lens}$. The relationship between the impact parameter
and $\theta$ is
\begin{equation}
\tan \theta = \frac{y_{\rm lens}}{x_e} \,.
\end{equation}
The distance between the center of the
globular cluster and the background star is given by $r_{\rm star}$.

%%%%% Figure %%%%%
%%%%%%%%%%

\subsection{Typical example -- Omega Centauri}
As an example for our study, one well-known globular is selected and verified.
Omega Centauri is the most massive globular cluster of the Milky Way.
The existence of an IMBH in Omega Centauri has been discussed 
from both theoretical and observational perspectives
~\citep{Noyola2010,Haggard2013,Baumgardt2019}.
The effect of the GL will be calculated using the known parameters of
Omega Centauri: $M_{\rm tot}=4 \times 10^6 M_{\odot}, r_{\rm lens}=82
 ~[\mbox{lyr}], x_e=1.56 \times 10^4  ~[\mbox{lyr}]$. For this calculation, $r_{\rm star}$ is
fixed at $5 \times 10^5~[\mbox{lyr}]$~\citep{vandeVen2006,DSouza2013}.

\begin{figure}
  \begin{center}
    \FigureFile(60mm,41mm){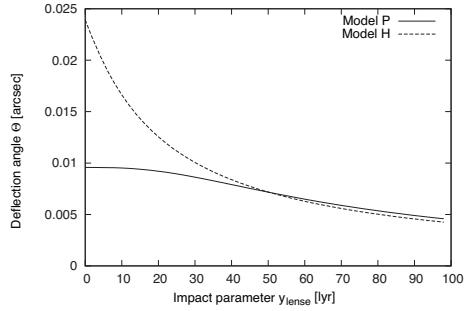}
  \end{center}
  \caption{The deflection angle by the Plummer model (Model P) and the Hernquist model (Model H) as a function of impact parameter.}
\label{fig:OmegaCentauri-P_H}
\end{figure}

When the cluster excludes the IMBH, the mass distribution is
given by the Plummer model or the Hernquist model. The deflection angle is 
shown in Fig.~\ref{fig:OmegaCentauri-P_H}. 
For the case of the Hernquist model (hereafter,
model H), 
the deflection angle increases sharply as the impact parameter decreases.
Conversely, for the case of the Plummer model (hereafter,
model P), the deflection angle approaches a constant value
at $y_{\rm lens}=0$.
This is due to the difference in gravitational potential between models:
for model P, the potential flattens at the center, but for model H, the potential sharpens there.

%%%%% Figure %%%%%
%%%%%%%%%%

\begin{figure}
  \begin{center}
    \FigureFile(60mm,41mm){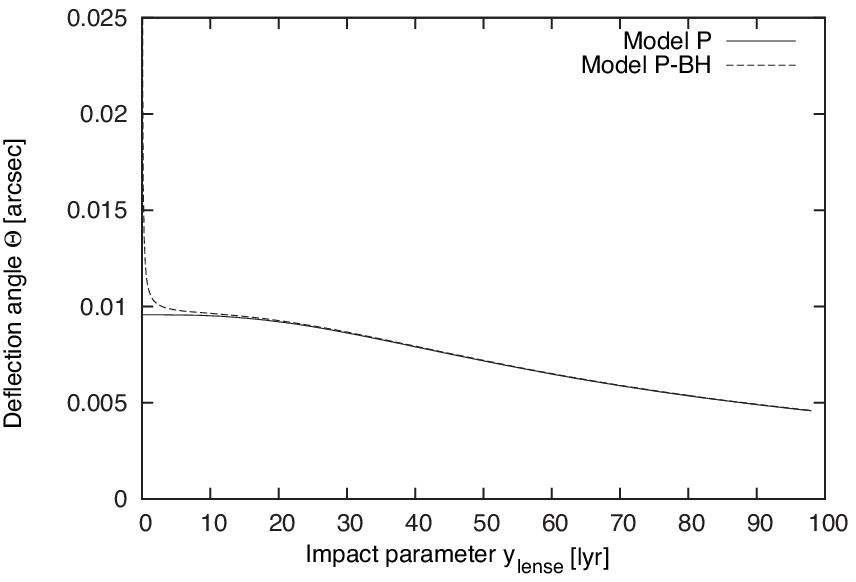}
    \FigureFile(60mm,41mm){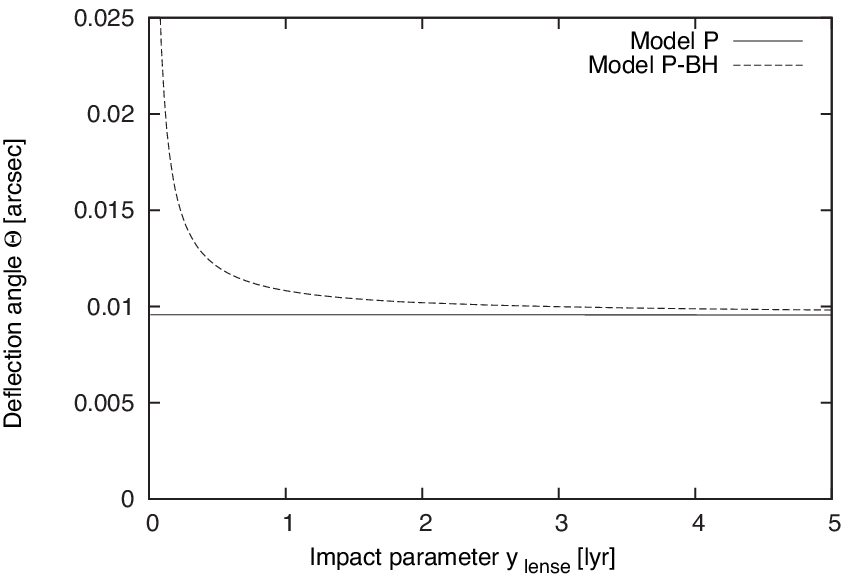}
  \end{center}
  \caption{Comparison of deflection angle with and without IMBH in the Plummer model. 
The figure on the right is an enlarged view of $y_{\rm lens}$ from the figure on the left.}
\label{fig:OmegaCentauri-P-BH}
\end{figure}

\begin{figure}
  \begin{center}
    \FigureFile(60mm,41mm){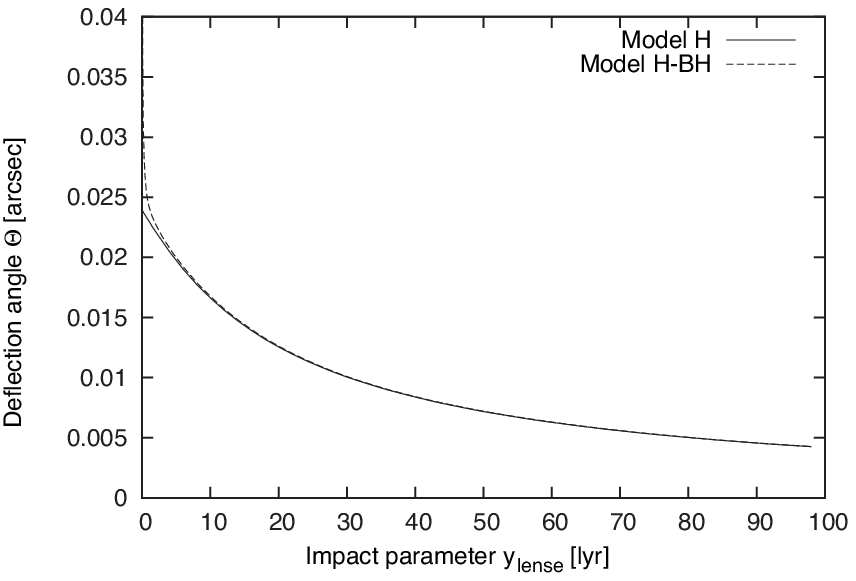}
    \FigureFile(60mm,41mm){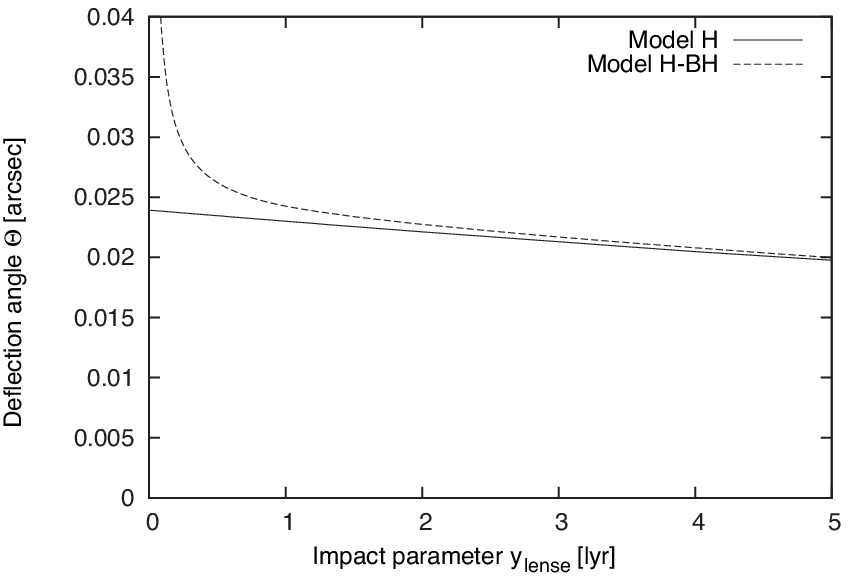}
  \end{center}
  \caption{Comparison of deflection angle with and without IMBH in the Hernquist model. 
The figure on the right is an enlarged view of $y_{\rm lens}$ from the figure on the left.}
\label{fig:OmegaCentauri-H-BH}
\end{figure}

When the cluster includes an IMBH, the deflection angle dramatically
changes at small impact parameters. For the case of the Plummer model
with an IMBH (hereafter, model P-BH), the difference of angle becomes
significant at $y_{\rm lens} \le 20 [\mbox{lyr}]$ (Fig.~\ref{fig:OmegaCentauri-P-BH}). 
Although
the boundary of Bahcall-Wolf cusp is fixed at $1 [\mbox{lyr}]$,
the trajectory of light is affected well outside this radius. Due to the effect of the IMBH, the
angle diverges at $r \simeq 0$. For the case of the Hernquist model
with an IMBH (hereafter, model H-BH),
although the effect of the IMBH (or cusp) appears at small impact parameters,
it is not as clear as in the case of model P-BH
 (Fig.~\ref{fig:OmegaCentauri-H-BH}).
Perhaps this result is caused by H-BH's less extreme potential slope near the center.

%%%%% Figure %%%%%
%%%%%%%%%%

%%%%% Figure %%%%%
%%%%%%%%%%

\subsection{Dependence on Parameters}
\label{sec:parameters}
\begin{figure}
  \begin{center}
    \FigureFile(130mm,97mm){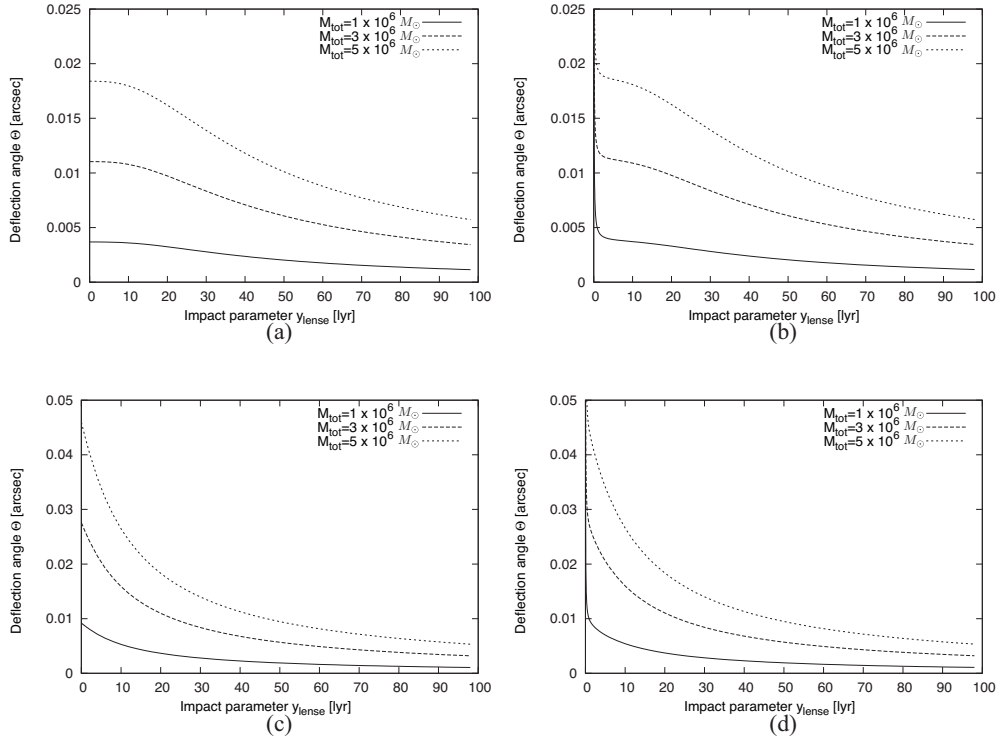}
  \end{center}
  \caption{Dependence of the deflection angle on the total mass $M_{tot}$ for
(a) Plummer model, (b) Plummer model with IMBH, (c) Hernquist model, and (d) Hernquist model with IMBH.}
\label{fig:Mtot}
\end{figure}

\begin{figure}
  \begin{center}
    \FigureFile(130mm,97mm){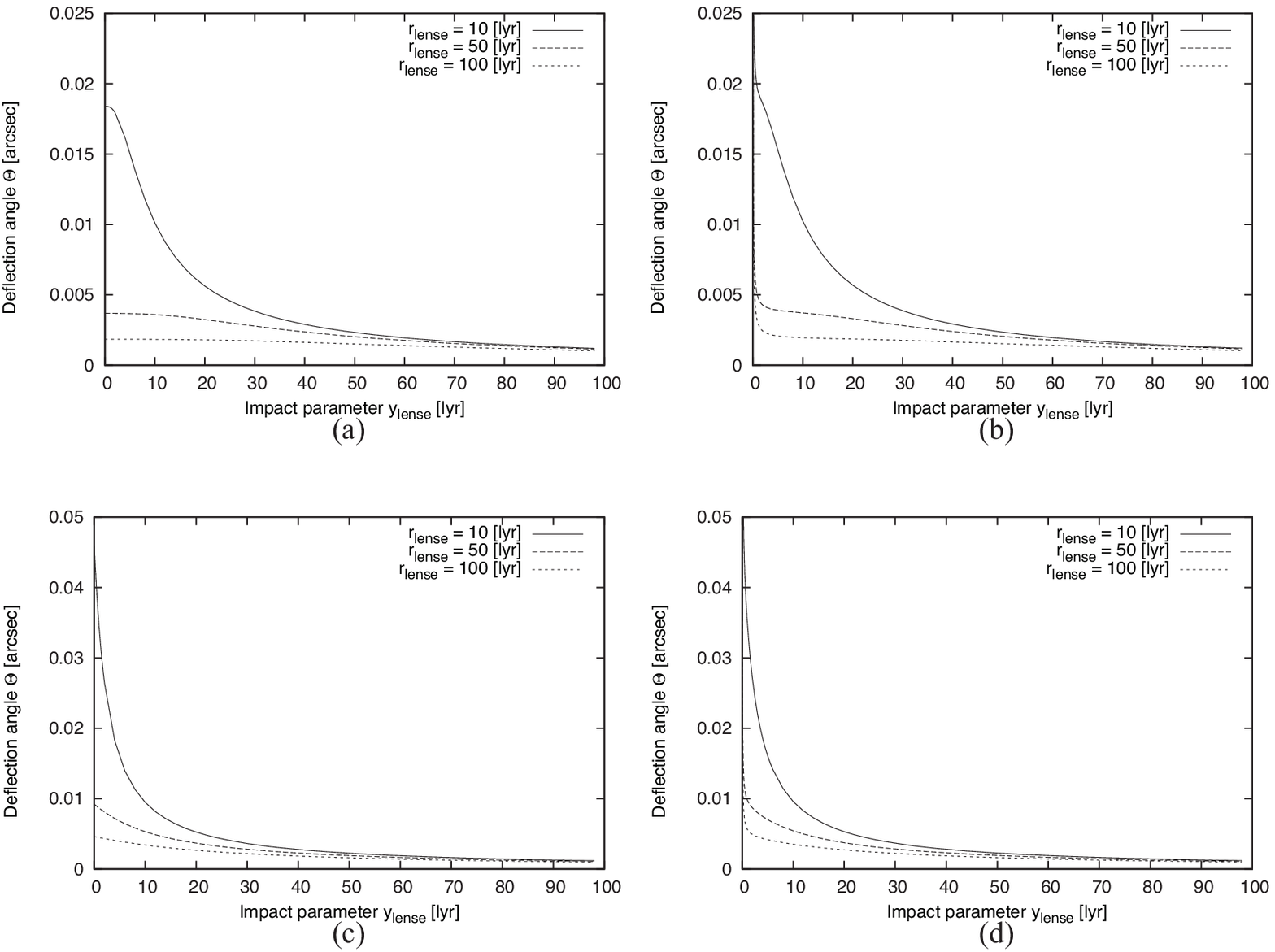}
  \end{center}
  \caption{Dependence of the deflection angle on the radius of the lensing object $r_{lens}$ for
(a) Plummer model, (b) Plummer model with IMBH, (c) Hernquist model, and (d) Hernquist model with IMBH.}
\label{fig:rlens}
\end{figure}

We investigate the effect of changing the parameters on the deflection angle.
The standard
values of the parameters are as follows:
\begin{itemize}
 \item Total mass of lensing object: $M_{\rm tot} = 10^6 M_{\odot}$
 \item Radius of lensing object: $r_{\rm lens}=50 [\mbox{lyr}]$
 \item Distance from the Earth: $x_e = 5 \times 10^4 [\mbox{lyr}]$
 \item Distance between the center of the globular cluster and the background object: \\
  $r_{\rm star} = 5 \times 10^5 [\mbox{lyr}]$
\end{itemize}
The above four parameters are then changed one by one, with the following results:

Fig.~\ref{fig:Mtot} shows the dependence of the deflection angle on total mass $M_{\rm tot}$.
As $M_{\rm tot}$ increases, the angle increases because gravity strengthens.
Fig.~\ref{fig:rlens} shows the dependence of the deflection angle on the radius  of the lensing object $r_{\rm lens}$. 
The lensing effect increases as the mass is concentrated in a narrower area.

%%%%% Figure %%%%%
%%%%%%%%%%

%%%%% Figure %%%%%
%%%%%%%%%%

\begin{figure}
  \begin{center}
    \FigureFile(130mm,97mm){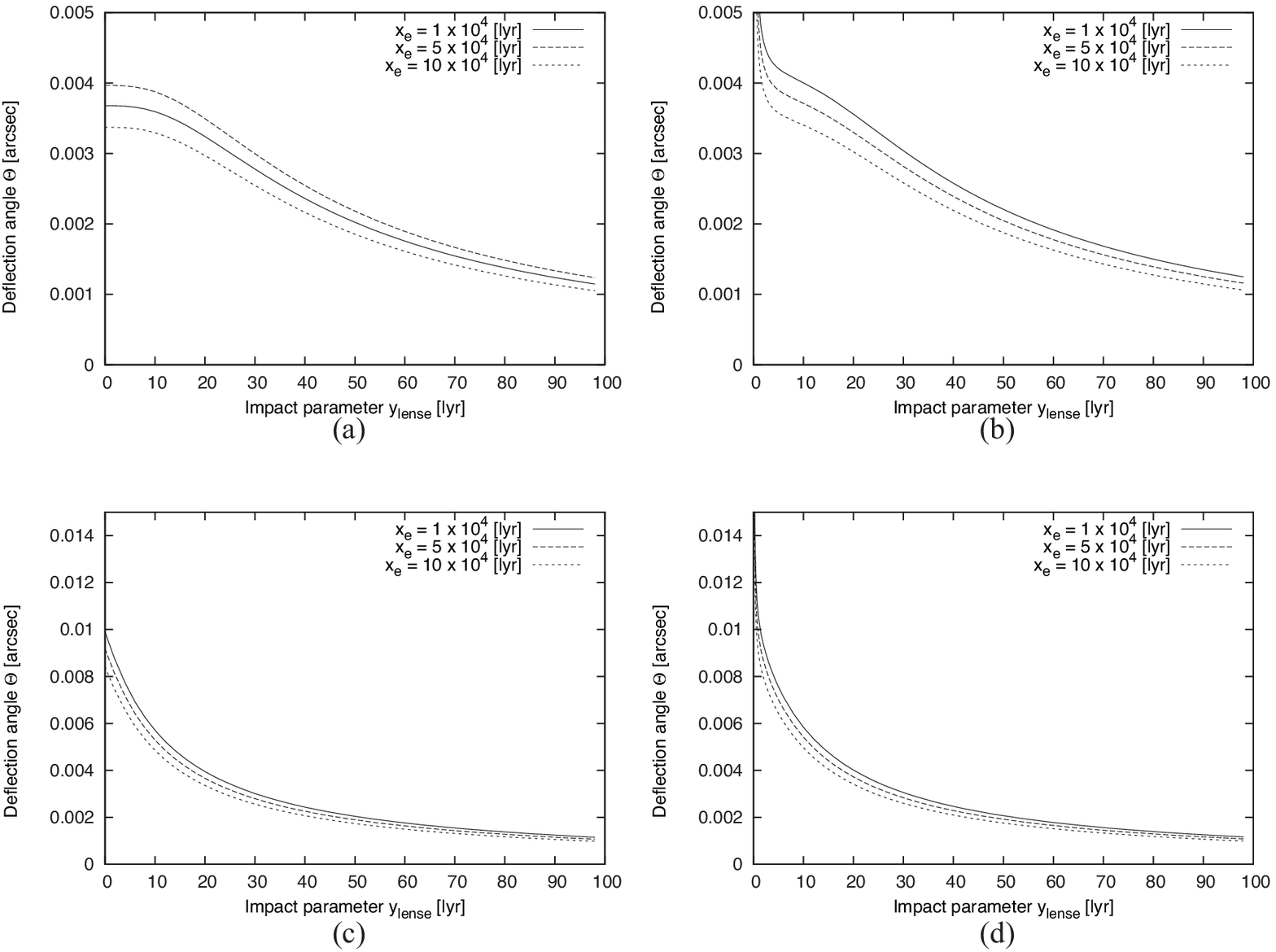}
  \end{center}
  \caption{Dependence of the deflection angle on distance between globular cluster and the Earth
$x_e$ for 
(a) Plummer model, (b) Plummer model with IMBH, (c) Hernquist model, and (d) Hernquist model with IMBH.}
 \label{fig:xe}
\end{figure}

We also show the dependence on the distance between the globular cluster and the Earth. The angle 
$\theta$ changes as the distance between the Earth and the globular
cluster changes. Fig.~\ref{fig:xe} shows the $\theta$ dependence of 
the deflection angle.
Since we are considering globular clusters belonging to the Milky Way,
the upper limit of the distance is set to $10^5 [\mbox{lyr}]$. 
The distance dependence of the angle therefore seems to be insignificant from the figure.

\begin{figure}
  \begin{center}
    \FigureFile(130mm,97mm){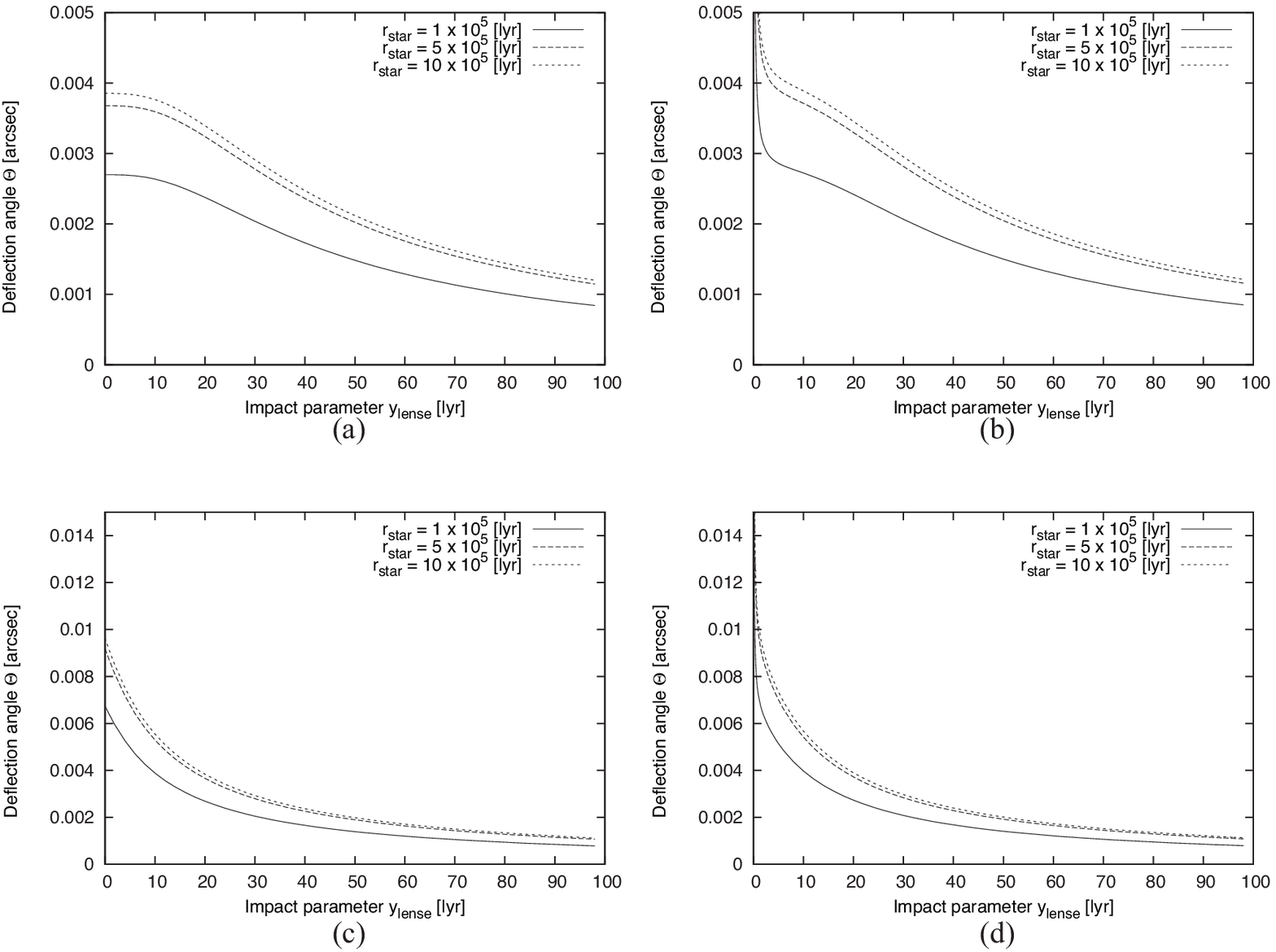}
  \end{center}
  \caption{Dependence of the deflection angle on distance to background object 
$r_{star}$ for 
(a) Plummer model, (b) Plummer model with IMBH, (c) Hernquist model, and (d) Hernquist model with IMBH.}
 \label{fig:rstar}
\end{figure}

Finally, we consider the dependence on distance to the background object
(Fig.~\ref{fig:rstar}).  
As the distance to the background object increases, the angle approaches a
limiting curve that, except in Model P, rises steeply at low impact parameters.

%%%%% Figure %%%%%
%%%%%%%%%%

%%%%% Figure %%%%%
%%%%%%%%%%

\subsection{Observability of Gravitational Lensing Effects}
\begin{figure}
  \begin{center}
    \FigureFile(80mm,26mm){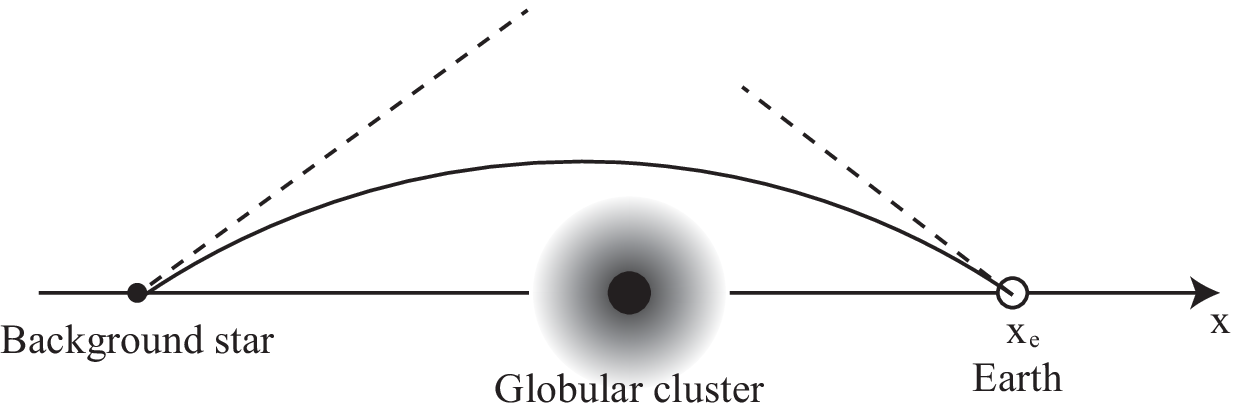}
  \end{center}
  \caption{Schematic of the ideal case in which the Earth, globular cluster, and background star
are aligned. Dashed tangential lines indicate the trajectory at the Earth and 
background star, respectively.
The separation angle becomes twice the refraction angle.}
\label{fig:ideal}
\end{figure}

Finding the trajectories of two light beams reaching the Earth from the same
astronomical object generally requires solving a boundary value problem.
The trajectory that connects the Earth and astronomical object is determined.
As a simple case, we consider a situation where the Earth, lensing
object, and background object are aligned (Fig.~\ref{fig:ideal}). 
In this situation, the separation angle of the two trajectories due to
GL can be calculated easily: 
it becomes $2 \Theta$, or twice the refraction angle.
If the separation angle is large enough, multiple celestial images
will be observed. Even if the separation angle is small, the effect
of GL will increase the brightness of the background
object.

%%%%% Figure %%%%%
%%%%%%%%%%

The Earth, lensing object, and background object are aligned
when the initial emission angle and the refraction angle are equal.
Therefore, for each model, the situation where the initial
emission angle and the refraction angle coincide has been investigated.
For analysis, we consider the parameters given by the case of
Omega Centauri. The results are shown in Table~\ref{tab:angle}.

\begin{table}
\caption{Impact factor $y_{\rm lens}$ when initial emission angle equals refraction angle $\Theta$}
\label{tab:angle}
  \begin{center}
    \begin{tabular}{ccc} \hline
Models & $y_{\rm lens} [\times 10^3 ~\mbox{lyr}]$ &
 $\Theta [\times 10^{-2} ~\mbox{arcsec}]$ \\  \hline
Model P & $0.724$ & $0.957$ \\
Model P-BH & $10.1$ & $13.1$ \\
Model H & $1.81$ & $2.39$ \\
Model H-BH & $10.8$ & $14.2$ \\ \hline
    \end{tabular} 
  \end{center}
\end{table}

From Table~\ref{tab:angle},
the presence of a black hole affects the separation angle
by about $10 \times 10^{-2} [\mbox{arcsec}]$. 

We have investigated the effects of other parameters also.
The value that maximizes the separation angle is selected
within the range of the parameters treated in Section
\ref{sec:parameters}:
\begin{itemize}
 \item $M_{\rm tot} = 5 \times 10^6 M_{\odot}$, 
 \item $r_{\rm lens} = 10$ [lyr],
 \item $x_e = 1 \times 10^4$ [lyr],
 \item $r_{\rm star} = 10 \times 10^5$ [lyr].
\end{itemize}
With this choice of parameters, the separation angle of the
H-BH model takes its maximum value, about $0.76$ [arcsec].
The smallness of this angle makes it difficult to observe the lensing
phenomenon from the ground due to complications
with the Earth’s atmosphere.

\subsection{Case of dark globular cluster}
As we mentioned in section 3, globular clusters in the Milky Way have   been cataloged
~\citep{Harris1996,Hilker2020}.
Our calculations indicate that the GL effect 
is small even in the case of Omega Centauri. 
It is unlikely that a new globular cluster will be found near the Earth.
It takes enough time for known globular clusters
to be close to Earth. Here we
consider a different situation.

Suppose that cold dark matter can condense into something like a globular cluster
~\citep{Carr1987}. 
Such an object would not emit light by itself and would not be included 
in the existing catalogs of globular cluster.
If such a ``dark globular cluster'' exists and is closer 
to the Earth than other globular clusters, how much
of a GL effect would it produce?

Here we consider dark globular clusters that
can be treated by the Hernquist model, with and without IMBHs. We fix two parameters:
\begin{itemize}
 \item $x_e = 5 \times 10^3$ [lyr]
 \item $r_{\rm star} = 5 \times 10^5$ [lyr]
\end{itemize}
%

%%%%% Figure %%%%%
\begin{figure}
  \begin{center}
    \FigureFile(60mm,41mm){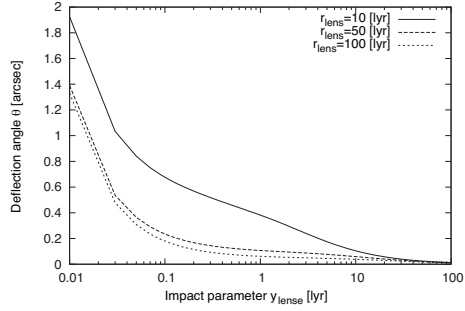}
  \end{center}
  \caption{Comparison of deflection angle when radius of lensing object is changed.}
\label{fig:Dark-M1M}
\end{figure}
%%%%%%%%%%

%%%%% Figure %%%%%
\begin{figure}
  \begin{center}
    \FigureFile(60mm,41mm){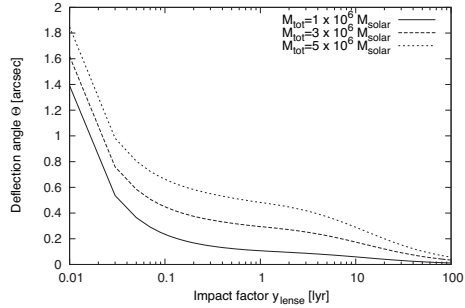}
  \end{center}
  \caption{Comparison of deflection angle when total mass of globular cluster is changed.}
\label{fig:Dark-SR50}
\end{figure}
%%%%%%%%%%

Fig.~\ref{fig:Dark-M1M} shows the deflection angle
when the radius of the lensing object is changed.
In this figure, the total mass of the globular cluster
is fixed at $M_{\rm tot} = 10^6 M_{\odot}$.
Fig.~\ref{fig:Dark-SR50} shows the deflection angle
when the total mass of the lensing object is changed.
In this figure, the radius of the globular cluster is fixed at
$r_{\rm lens} =50 ~[\mbox{lyr}]$,

In these models, 
the deflection angle can be up to about 2 [arcsec]. 
If a dark globular cluster containing an IMBH is within 
a few thousand light years from the Earth, it may
be detected by the GL effect 
in future observations.

%%%%%%%%%%%%%%%%%%%%%%%%%%%%%%%%%%%%%%%%
%%% Conclusion
%%%%%%%%%%%%%%%%%%%%%%%%%%%%%%%%%%%%%%%%

\section{Conclusion}
We considered a method to determine the presence of an IMBH in
a globular cluster by examining the GL effect. Under the assumption of spherical
symmetry, we considered the mass distribution of globular clusters 
with and without IMBHs, and then calculated
the separation angle of light trajectories based on this.
(Note that, in this study, it was assumed that the trajectories could
pass through the interior of the globular cluster, but such a pass-through might be impossible if the globular cluster is
extremely dense. In that case, background objects could not be seen.)

A catalog of globular clusters in the Milky Way Galaxy
has been published~\citep{Harris1996,Hilker2020}.
Typical parameters, based on those listed in this catalog,
served as input to see how
strong GL effect a cluster can produce. For Omega Centauri, 
a dense globular
cluster that is the most massive one belonging to the Milky Way, 
the separation angle of the GL effect
caused by the globular cluster with an IMBH was found to be
on the order of sub-milliarcseconds.
For other globular clusters belonging to the Milky Way,
the separation angle would be even smaller.

Observing multiple images may be difficult due to the small separation 
angle of the GL.
The brightening effect caused by GL may be more easily observed.
Light that travels through multiple trajectories from
background objects can be concentrated onto the Earth by the GL
effect, which causes sources to appear brighter than
they are in reality. 
Systematic observations of the GL effect
on galaxies have been made in the past~\citep{Tyson1990,Dahle2002,Kubo2009,Jaelani2020}.
We are considering similar surveys for globular clusters.

This method has been applied to MACHO observations
with small separation angles
\citep{MACHO}. If a single observation reveals that the
brightening effect of the globular cluster due to the GL
effect is quite strong, the density distribution
of the globular cluster may be high, suggesting the
presence of an IMBH. 
By observing the change in the brightening rate of a globular
cluster as it passes between background objects and the Earth,
it may be possible to detect the presence or absence of an IMBH,
but such continuous observations would require a more extensive observation program than
the MACHO Project.
Quantitative evaluation of brightening due to the
GL effect will be derived
in future work.

\bigskip

The authors would like to thank Shigeyuki Karino, Takahiko Matsubara, Kouji Nakamura, Hiroyuki Nakano,
Hisaaki Shinkai, Kousuke Sumiyoshi for their useful discussion at 32nd RIRONKON symposium
(Dec. 25-27, 2019, National Astronomical Observatory, Japan).
We would like to thank Editage (www.editage.com) for English language editing.

%\appendix
%\section{Method of .....}
%
%\section{Approximation of ...}
%
%\section*{Complete data}

%%%
% See the manual for the detail.
%%%

\end{document}